\documentclass[%
reprint, showpacs,
 amsmath,amssymb,
 aps,
 prb,
]{revtex4-1}

\usepackage{graphicx}
\usepackage{dcolumn}
\usepackage{bm}
\usepackage{subfigure}
\usepackage{color}

\newcommand{\ve}[1]{{\bf{#1}}}
\newcommand{\ma}[1]{\underline{\underline{#1}}}

\def\be#1\ee{\begin{equation}#1\end{equation}}
\def\bea#1\eea{\begin{align}#1\end{align}}
\def\bse#1\ese{\begin{subequations}#1\end{subequations}}

\def\1#1{{\hat{{\boldsymbol{#1}}}}}                                 		
\def\2#1{\hat{#1}}                                              		   		
\def\3#1{{\mathbf{#1}}}                                             	   		
\def\4#1{{\boldsymbol{#1}}}                                            		
\def\5#1{{\mathcal#1}}                                                            		
\def\6#1{\bar{#1}}                                                           		
\def\7#1{{{#1}}}                                    		
\def\8#1{\widetilde{#1}}
\def\9#1{\check{#1}}
\def\+#1{{\overset{{\scriptscriptstyle +}}{#1}{}}}                  		
\def\b+#1{{\overset{{\scriptscriptstyle +}}{\mathbf{#1}}{}}}        	
\def\g+#1{{\overset{{\scriptscriptstyle +}}{\boldsymbol{#1}}{}}}    

\definecolor{dark-green}{rgb}{0.278,0.7,0.4}                    

\definecolor{my-brown}{rgb}{0.69,0.247,0.13}                    

\definecolor{my-purple}{rgb}{0.47,0.12,0.46}                    

\definecolor{my-greenblue}{rgb}{0.129,0.313,0.419}              

\definecolor{my-orange}{rgb}{1,0.5,0.25}                        

\definecolor{my-red}{rgb}{0.745,0,0.2117}                       

\definecolor{my-gray}{rgb}{0.5,0.5,0.5}                         

\definecolor{my-dark-blue}{rgb}{0.1,0.1,0.7}                    

\definecolor{my-indigo}{rgb}{0.29,0.0,0.51}                    

\begin{document}

\preprint{APS/123-QED}



\title{Space-Time Modulated Loaded-Wire Metagratings for Magnetless Nonreciprocity and Near-Complete Frequency Conversion}

\author{Yakir Hadad}
\email{hadady@eng.tau.ac.il}
\affiliation{School of Electrical Engineering, Tel-Aviv University, Ramat-Aviv, Tel-Aviv, Israel, 69978}

\author{Dimitrios Sounas}
\affiliation{Department of Electrical and Computer Engineering, Wayne State University, Detroit, MI 48202, USA}

\date{\today}

\begin{abstract}
In recent years a significant progress has been made in the development of magnet-less nonreciprocity using space-time modulation, both in electromagnetics and acoustics.
This approach has so far resulted in a plethora of non-reciprocal devices, such as isolators and circulators, over different parts of the spectrum, for guided waves. On the other hand, very little work has been performed on non-reciprocal devices for waves propagating in free space, which can also have many practical applications. For example, it was shown theoretically that non-reciprocal scattering by a metasurface can be obtained if the surface-impedance operator is continuously modulated in space and time. %
However, the main challenge in the realization of such a metasurface is due to the high complexity required to modulate in space and time many sub-wavelength unit-cells of which the metasurface consists.
In this paper we show that spatiotemporally modulated metagratings can lead to strong nonreciprocal responses, despite the fact that they are based on electrically-large unit cells. We specifically focus on wire metagratings loaded with time-modulated capacitances. We use the discrete-dipole-approximation and an ad-hoc generalization of the theory of polarizability for time-modulated particles, and demonstrate an effective nonreciprocal anomalous reflection (diffraction) with an efficient frequency conversion. Thus, our work opens a venue towards a practical design and implementation of highly non-reciprocal magnet-less metasurfaces in electromagnetics and acoustics.
\end{abstract}

\pacs{Valid PACS appear here}
\maketitle



\section{Introduction} 
In recent years substantial efforts have been devoted towards the development of magnet-less non-reciprocal devices  in electromagnetics as well as in acoustics \cite{Tanaka1965, Ayasli1989, Gallo2001, Kodera2011, Sounas2013, Wang2012, Popa2012, Manipatruni2009,  Mazor2019_2, Fleury2014}. In particular, space-time modulated structures have been explored theoretically and experimentally for this purpose \cite{Qin2014, Reiskarimian2016, Yu2009, Lira2012, Qin2016, Sounas2013_2, Estep2014}. These designs are based on the creation of a synthetic sense of motion \cite{Mazor2019} that fundamentally enables the breach of time-reversal symmetry \emph{with respect to the guiding-wave sub-system}.
While the major part of these efforts has been dedicated to breaking reciprocity in guiding wave structures, some important focus has been also aimed at the violation of reciprocity in scattering and radiation \cite{Hadad2015, Hadad2016, Taravati2017, Taravati2017_2}.
For example in \cite{Hadad2016} a space-time modulated travelling wave antenna with different radiation patterns in transmit and receive mode was studied theoretically and experimentally. In that antenna, a few  voltage varying capacitors were used to establish the required modulation. The underlying idea of this design is to amplify the effect of the spatiotemporal modulation by taking advantage of asymmetric interband transitions between different space-time harmonics of  a mode inside and outside the light-cone. Thus, in a sense, the natural sharp filtering property of the light cone has been used to enhance the nonreciprocal effect in this low-Q leaky wave antenna system.
In another work \cite{Hadad2015}, a metasurface that consists of a space-time modulated impedance operator has been explored theoretically. In that proposal non-reciprocal Wood's anomaly has been achieved through a careful design of the space-time modulation of the surface characteristics.
However, unlike the antenna in Ref.~\cite{Hadad2016} which required a small number of modulation actuators, for the emulation of the effective surface impedance  in \cite{Hadad2015}, an unrealistically, extremely complex  configuration of voltage varying capacitors and dense wiring system is required to achieve the many different modulation  regions.

A possible venue to overcome this issue may be based on using metagratings which are metasurfaces with electrically large unit cells. Metagratings \cite{Chalabi2017, Radi2017, Epstein2017, Rabinovich2018, Popov2018, Popov2019, Popov2019b, Epstein2018, Diaz-Rubio2017, Taravati2019}, a close relative of frequency selective surfaces and diffraction gratings, have been explored recently as another means to overcome the challenge of perfect anomalous reflection and refraction by an electrically thin metasurface \cite{Yu2011, Monticone2013, Li2014, Pfeiffer2013, Selvanayagam2013, Selvanayagam2013_2, Pfeiffer2013_2, Sievenpiper2002, Hasman2003, Lin2014, Wong2014, Kim2014, Epstein2016, Epstein2014, Epstein2016_2, Dorrah2018, Sounas_Estakhri2016}.
As opposed to the traditional metasurface approach in which electrically deep subwavelength unit cells are involved, the unit cell of the metagrating is ${\cal O}(\lambda/2)$ or more and thus fundamentally gives rise to propagating diffraction orders. Then, a proper design of the metagrating's unit-cell enables a significant control over the balance of the various propagating diffraction orders, and in particular makes it possible to nullify the zero order harmonic which corresponds to regular reflection and refraction.
Moreover, recently, an interesting proposal for a dielectric slab metagrating with space-time modulated refractive index has been explored demonstrating a peculiar relations between the space-time harmonics in this system \cite{Taravati2019}. In particular, non-reciprocal scattering has been theoretically demonstrated.
Nonetheless, a realistic implementation of space-time modulated refractive index is challenging and requires either nonlinear media with significant pump power or the use of effective medium, e.g., using a dense array of unit-cells that are loaded by voltage-varying capacitors.
Thus, unfortunately raising again the challenge of realizability.

Here, inspired by the recent progress in metagratings \cite{Chalabi2017, Radi2017, Epstein2017, Rabinovich2018, Popov2018, Popov2019, Popov2019b, Epstein2018, Diaz-Rubio2017, Taravati2019}, we propose a solution to this problem and design efficient non-reciprocal metasurfaces by using only a minimal  number of \emph{three} distinct modulation regions and small modulation parameters. Specifically, we develop a rigourous theory for  metagratings with time-modulated resonant elements. Using this approach we explore analytically a two-dimensional  lattice of resonant, capacitively loaded, wires that are subject to space-time modulation.
%
%
%
To that end we first develop a generalization of polarizability theory for time-modulated particles, and later exploit it together with the discrete-dipole approximation \cite{Draine1994, Tretyakov_Book,  Hadad2013, Hadad2010, Mazor2012, Mazor2015} and the proper Green's function to explore analytically the scattering from an infinite space-time modulated metagratings. We particularly show that non-reciprocal anomalous reflection can be achieved with nearly perfect coupling efficiency between the incident wave and the desired scattered wave. As opposed to anomalous reflection obtained by a stationary metagratings, here, the anomalous reflection process involves also an efficient frequency conversion process.

\section{The generalized polarizability of time-modulated loaded wire}
\begin{figure}[htbp]
  \centering
  \includegraphics[width=81mm]{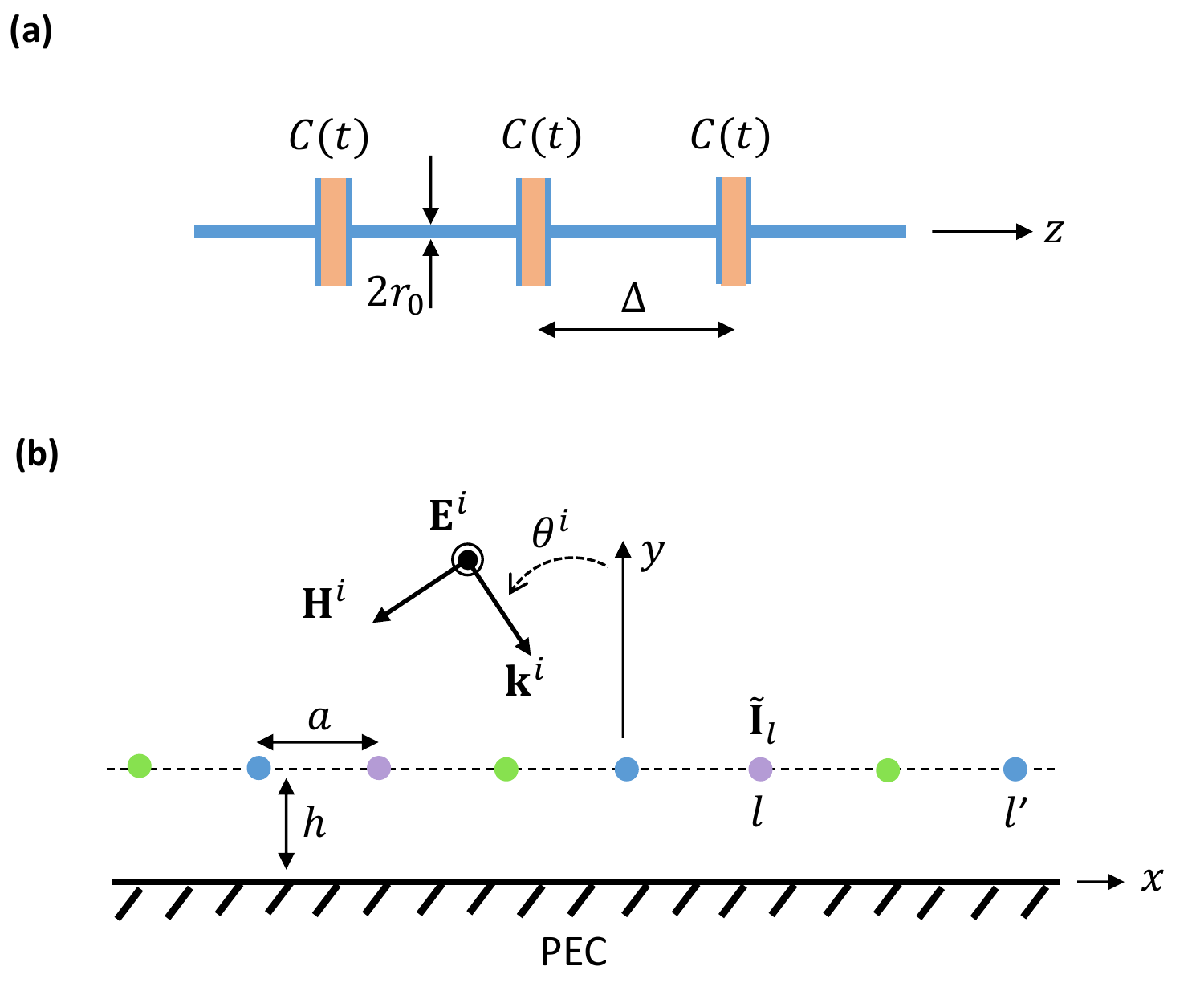}
  \caption{Illustration of the problem. (a) Temporally modulated capacitively loaded wire, co-aligned with the $z$-axis. (b) A space time modulated metagratings wire array above a perfect electrically conducting plate. Here, the spatial periodicity is $N=3$, the minimal required to achieve non-reciprocity by synthetic motion.}
  \label{fig1}
\end{figure}

Consider a periodically loaded perfect electrically conducting (PEC) wire that is co-aligned with the $z$-axis as illustrated in Fig.~\ref{fig1}(a). The wire radius is $r_0$, the loading impedance is $Z_L$, and the loading periodicity is small on the wavelength, i.e., $\Delta\ll\lambda$.  Now let $\ve{E}(\omega)=\hat{z}E(\omega)$ be the electric field at the wire location but in the absence of the wire itself. Then, the induced current on the wire is given by \cite{Tretyakov_Book}
\begin{equation}\label{Induced curr FD}
I(\omega)=\alpha(\omega)E(\omega)
\end{equation}
with the effective susceptibility\cite{Comm} $\alpha$,
\begin{equation}
\alpha^{-1}(\omega)=\alpha^{-1}_0(\omega) + \frac{Z_L}{\Delta}
\end{equation}
where
\begin{equation}\label{alp unloaded}
\alpha_0^{-1}(\omega)=\frac{\eta k}{4}H_0^{(2)}\left(k r_0\right)
\end{equation}
is the susceptibility  of an unloaded PEC  wire. In Eq.~(\ref{alp unloaded}), $\eta=120\pi\Omega$ and $k=\omega/c$  are the free space impedance and   wavenumber, respectively. Also, $\omega$ is the radial frequency, $c$ is the speed of light in vacuum, and $H_0^{(2)}$ is zero order Hankel function of the second kind.
To simplify subsequent notations, in the following we use
\begin{equation}\label{gamma def}
\gamma(\omega)=\frac{1}{\alpha(\omega)}.
\end{equation}
Then, for \emph{capacitive} periodic loading we can write
\begin{equation}\label{gamma loaded}
\gamma(\omega)=\gamma_0(\omega) + \frac{1}{j\omega C \Delta}
\end{equation}
where $C$ is the per-unit-length loading capacitance.

Up to this point the formulation has been strictly carried out in the frequency domain. However, in the following we shall introduce temporal modulation of the loading capacitor. We assume
\begin{equation}\label{Cap Time Mod}
\frac{1}{C(t)}=\frac{1}{C_0}\left[1 + m\cos(\omega_m t + \phi) \right],
\end{equation}
where $\omega_m$ is the modulation frequency, $m$ is the modulation index and $\phi$ is the modulation phase shift between the wires in the meta-grating structure.
To model the scattering process in this case  it is required to repeat the previous derivation, but in the time-domain. To that end we use  the convolution property of the Fourier transform on Eq.~(\ref{Induced curr FD}) with Eq.~(\ref{gamma def}). We have
\begin{equation}\label{E by conv}
{\cal E}(t)=\gamma(t)*{\cal I}(t)=\int_{-\infty}^{\infty}\gamma(\tau){\cal I}(t-\tau)d\tau
\end{equation}
where ${\cal E}(t), \gamma(t)$, and ${\cal I}(t)$ are the time domain counterparts of $E(\omega), \gamma(\omega)$, and $I(\omega)$, respectively, and $*$ denotes convolution as defined in Eq.~(\ref{E by conv}).
By explicitly expanding Eq.~(\ref{E by conv}) using Eq.~(\ref{gamma loaded}), we get
\begin{eqnarray}\label{E of t}
  {\cal E}(t) &=& \gamma_0(t)*{\cal I}(t)+\frac{1}{\Delta C(t)}{\cal F}^{-1}\left\{\frac{I(\omega)}{j\omega} \right\}(t) \\
   &=& \gamma_0(t)*{\cal I}(t) + \frac{1}{\Delta C(t)}\int_{-\infty}^t {\cal I}(\tau) d\tau
\end{eqnarray}
Here ${\cal F}^{-1}$ stand for the inverse Fourier transform. In the last  equality we used the identity ${\cal F}^{-1}\{I(\omega)/j\omega\} =  \int_{-\infty}^t {\cal I}(\tau)d\tau -  0.5I(\omega)|_{\omega=0}$, together with the fact $I(\omega)|_{\omega=0}=0$ (implying no DC currents in the problem).

Assuming that the exciting field is harmonic at frequency $\omega$, then, due to modulation at $\omega_m$,
the induced current and the electric field in the problem may be expressed by
\begin{equation}\label{Temp expan}
{\cal X}(t)=\frac{1}{2}\sum_{n=-\infty}^{\infty}{\cal X}_n e^{j\omega_n t} + c.c.
\end{equation}
where ${\cal X}$ stands for either ${\cal I}$ or ${\cal E}$, and  ${\cal X}_n$ denotes  the complex coefficient of n'th harmonic at  $\omega_n=\omega+n\omega_m$.
Next, in order to obtain a difference equation that relates between the different harmonics of the exciting electric field and the induced current, we apply Eq.~(\ref{Temp expan}) in Eq.~(\ref{E of t}). Beginning with the last term in Eq.~(\ref{E of t}), we find after a simple derivation,
\begin{eqnarray}\label{int of t}
  &&\frac{1}{C(t)}\int_{-\infty}^t  {\cal I}(\tau) d\tau = \frac{1}{2C_0} \sum_{n=-\infty}^{\infty}\!\! {e^{j\omega_n t}}\!\! \left[\frac{{\cal I}_n}{j\omega_n} + \right. \nonumber\\
   &&  \left.\frac{m}{2}e^{j\phi}\frac{{\cal I}_{n-1}}{j\omega_{n-1}}+ \frac{m}{2}e^{-j\phi}\frac{{\cal I}_{n+1}}{j\omega_{n+1}} \right]\!  + \! c.c.
\end{eqnarray}
Next, we evaluate the first term in Eq.~(\ref{E of t}). To that end we use the convolution identity
\begin{equation}
\gamma_0(t)*{\cal I}(t)={\cal F}^{-1} \left\{\gamma_0(\omega)I(\omega) \right\}
\end{equation}
with the Fourier transform of Eq.~(\ref{Temp expan}) (here, ${\cal X}={\cal I}$)
\begin{equation}
I(\omega)=\frac{1}{2}\sum_{n=-\infty}^{\infty}{\cal I}_n\delta(\omega-\omega_n)+c.c.
\end{equation}
Therefore leading to
\begin{equation}\label{gamma0 conv I}
\gamma_0(t)*{\cal I}(t) = \frac{1}{2}\sum_{n=-\infty}^{\infty} {\cal I}_n \gamma_0(\omega_n) e^{j\omega_n t} + c.c.
\end{equation}
Finally, by combining Eqs.~(\ref{gamma0 conv I}) and (\ref{int of t}) in Eq.~(\ref{E of t}) and with Eq.~(\ref{Temp expan}), we get
\begin{eqnarray}\label{En by In}
  && {\cal E}_n={\cal I}_n\gamma_0(\omega_n) + \nonumber\\
  &+& \frac{1}{\Delta C_0} \left[\frac{{\cal I}_n}{j\omega_n}+ \frac{m}{2}e^{j\phi}\frac{{\cal I}_{n-1}}{j\omega_{n-1}} + \frac{m}{2}e^{-j\phi}\frac{{\cal I}_{n+1}}{j\omega_{n+1}} \right].
\end{eqnarray}
Using Eq.~(\ref{En by In}) it is possible to define a ``generalized'' effective susceptibility $\alpha$ for time-modulated wires. In this case, $\alpha$ becomes a matrix that relates between the electric field and the induced currents in different harmonics. For example, if we choose to take into account only three harmonics, $n=0,\pm1$, that is at $\omega_0=\omega, \omega_{\pm1}=\omega\pm\omega_m$ we get
\begin{equation}
\tilde{\ve{I}}=\ma{\alpha}\tilde{\ve{E}}
\end{equation}
where $\tilde{\ve{E}}=[{\cal E}_{-1},{\cal E}_0,{\cal E}_1]^T$ and $\tilde{\ve{I}}=[{\cal I}_{-1},{\cal I}_0,{\cal I}_1]^T$, and
\begin{widetext}
\begin{equation}\label{generalized pol}
\ma{\alpha}^{-1}=\begin{bmatrix}
                    \gamma_0(\omega_{-1}) + \dfrac{1}{j\Delta C_0\omega_{-1}} & \dfrac{me^{-j\phi}}{2j\Delta C_0\omega_{0}} & 0 \\ \\
                    \dfrac{me^{j\phi}}{2j\Delta C_0\omega_{-1}} & \gamma_0(\omega_{0}) + \dfrac{1}{j\Delta C_0\omega_{0}} & \dfrac{me^{-j\phi}}{2j\Delta C_0\omega_{1}} \\ \\
                    0 & \dfrac{me^{j\phi}}{2j\Delta C_0\omega_{0}} & \gamma_0(\omega_{1}) + \dfrac{1}{j\Delta C_0\omega_{1}}
                  \end{bmatrix}
\end{equation}
\end{widetext}

\section{A Linear Array of Time-Modulated Capacitively Loaded Wires}\label{Linear Array}
Consider a linear array of capacitively loaded time-modulated wires, located at  $y=h$, above an infinite PEC plate that is placed on the $y=0$ plane. The inter-wire spacing is $a$ and the wires are loaded with capacitors $C(t)$, as in Eq.~(\ref{Cap Time Mod}), with
\begin{equation}
\phi=-\frac{2\pi l}{N}
\end{equation}
where $l\in \mathbb{Z}$ is the wire index and $N$ is the spatial periodicity.
For simplicity, in the following we shall assume that $N=3$, which is the minimal spatial periodicity that is required to achieve the effect of a synthetic motion, and thus, non-reciprocity.
An illustration of the structure in this case is shown in Fig.~\ref{fig1}(b).

\subsection{Formulation of the excitation dynamics}
The lattice is excited by an impinging plane wave
\begin{equation}\label{incident wave}
\ve{E}^{i}(\ve{r};\omega)=\hat{z}E^ie^{-j\ve{k}^{i}\cdot\ve{r}}
\end{equation}
where $\ve{r}=(x,y)$, $\ve{k}^{i}=(k^i_x,-k^i_y)$ where $k^i_x=k\sin\theta^i$ and $k^i_y=k\cos\theta^i$ and with $-90^\circ<\theta^i<90^\circ$ being the angle of incidence with respect to the normal $\hat{y}$. Clearly, $\theta^i$ is positive (negative) for waves with positive (negative) $k_x$ component of the wavenumber. See  Fig.~\ref{fig1}(b).

Using the generalized time-modulated susceptibility concept developed in the previous section, the equation of dynamics for the infinite array can be expressed as
\begin{eqnarray}\label{lattice dynamics}
  \ma{\alpha}_l^{-1}\tilde{\ve{I}}_l  &=& \tilde{\ve{E}}^{i}(\ve{r}_l)  \nonumber\\
   &+& \sum_{l'\neq l} \ma{G}(|\ve{r}_l-\ve{r}_{l'}|)\tilde{\ve{I}}_{l'} - \sum_{l'}\ma{G}(|\ve{r}_{l}-\ve{r}_{l'}^i|)\tilde{\ve{I}}_{l'}.
\end{eqnarray}
Here $\ma{\alpha}_l$ is the generalized susceptibility of the $l$-th wire, and
$\ma{G}$ is the two-dimensional free-space Green's function  evaluated at each of the intermodulation frequencies. Thus, in general,
\begin{equation}
\ma{G}=\mbox{diag}[...,G_{n-1},G_n,G_{n+1},...]
\end{equation}
where
\begin{equation}
G_n(|r-r'|)=\frac{-\eta k_n}{4}H_0^{(2)}(k_n|r-r'|)
\end{equation}
with $k_n=\omega_n/c$, and $\ve{r}=(x,y),\ve{r}'=(x',y')$ are the locations of the observer and the source, respectively.
In Eq.~(\ref{lattice dynamics}) $r_l=(h,la)$ is the locations of the $l$-th  wire on the $(x,y)$ plane, and $r_l^i=(-h,la)$ is the location of the \emph{image by the PEC plate} of the $l$-th wire. Thus,
in Eq.~(\ref{lattice dynamics}) we use image theory in order to replace the original problem with a new problem in which all the wires are located in free space. For this reason, the generalized susceptibility developed in the previous section for a wire in free-space is still valid.
However, it is essential to write correctly the incident field in this case. The latter consists of the  impinging wave  which is given in Eq.~(\ref{incident wave}), plus the reflected wave from the PEC plate in the absence of the wire array.
Specifically, on $\ve{r}_l=(al,h)$ for the case of $N=3$, and after taking into account only the three low order harmonics $n=0,\pm1$, the incident field in Eq.~(\ref{lattice dynamics}) reads,
\begin{equation}\label{incident field after image}
\tilde{\ve{E}}^{i}(\ve{r}_l)=[0,2jE^i\sin(k^i_yh)e^{-jk^i_xal},0]^T
\end{equation}
highlighting the fact that there are no intermodulation frequencies in the incident wave.

\subsection{The induced currents}
Due to Floquet-Bloch theorem, the $n$-th harmonic of the induced current on the $l$-th wire is given by
\begin{equation}
{\cal I}_{n,l}={\cal A}_n e^{-j\psi_nl}.
\end{equation}
where
\begin{equation}
\psi_n = k_x^i a+\frac{2\pi n}{3}.
\end{equation}
Consistent with previous approximation of week temporal modulation, we keep only the three fundamental harmonics, $n=0,\pm1$, and write,
\begin{equation}
\tilde{\ve{I}}_{l}=[{\cal I}_{-1,l},{\cal I}_{0,l},{\cal I}_{1,l}]^T.
\end{equation}
This solution anzats is applied in the infinite summations of Eq.~(\ref{lattice dynamics}). Beginning with the first summation in Eq.~(\ref{lattice dynamics}), for a specific harmonic $n$, and following similar derivation as in \cite{Tretyakov_Book}, we have
\begin{equation}
\sum_{l'\neq l}G_n(|\ve{r}_l-\ve{r}_{l'}|){\cal I}_{n,l'}=2{\cal A}_n e^{-j\psi_n l} \left(-\frac{\eta k_n}{4}\right) S_1^{(n)}
\end{equation}
where
\begin{eqnarray}
 && S^{(n)}_1 =  \sum_{s=1}^{\infty}H_0^{(2)}(k_n a s)\cos(\psi_ns)
   = \frac{1}{\beta_{0n} a} - \frac{1}{2} + \nonumber\\
   &&\frac{j}{\pi} \left[\ln\left(\frac{k_n a}{4\pi} \right) + \gamma +\frac{1}{2}\sum_{m\neq0}\left(\frac{-2\pi j}{\beta_{mn} a} -\frac{1}{|m|} \right) \right]
\end{eqnarray}
and with
\begin{equation}\label{betam}
\beta_{mn}  = \sqrt{k_n^2 - \left(\frac{2\pi m}{a} + \frac{\psi_n}{a}\right)^2}, \quad \Im\{\beta_{mn}\}\le 0.
\end{equation}
The second summation in Eq.~(\ref{lattice dynamics}) is treated using a conventional Poisson summation as in \cite{Tretyakov_Book}.
\begin{eqnarray}
 \sum_{l'}G_n(|\ve{r}_l-\ve{r}_{l'}^i|){\cal I}_{n,l'} =
 {\cal A}_n e^{-j\psi_nl}\left(-\frac{\eta k_n}{4} \right) S_2^{(n)}
\end{eqnarray}
where
\begin{eqnarray}
  S^{(n)}_2 &=& \sum_{s=-\infty}^{\infty} H_0^{(2)}(k_n\sqrt{|s|^2a^2 + 4h^2})e^{j\psi_n s}\nonumber \\
   &=& \frac{2}{a}\sum_{m=-\infty}^{\infty}\frac{e^{-j2\beta_{mn} h}}{\beta_{mn}} \\
\end{eqnarray}
with $\beta_{mn}$ as defined in Eq.~(\ref{betam}).
Once the summations are evaluated, using Eq.~(\ref{lattice dynamics}) it is possible to write an equation for the unknown excitation amplitudes ${\cal A}_n$. Consistent with previous notations we denote
\begin{equation}
\tilde{\ve{A}}=[...,{\cal A}_{n-1},{\cal A}_n,{\cal A}_{n+1},...]^T.
\end{equation}
Taking into account only three harmonics $n=0,\pm1$, we end up with solving the following linear equation for the excitation amplitudes at each harmonics,
\begin{equation}\label{Eq A}
(\ma{M}-\ma{R})\tilde{\ve{A}}=\tilde{\ve{E}}^{i}(\ve{r}_0)
\end{equation}
with
\begin{widetext}
\begin{equation}\label{Exc Amp}
\ma{M}=\begin{bmatrix}
    \gamma_0(\omega_{-1}) + \dfrac{1}{j\Delta C_0\omega_{-1}}  & \dfrac{m}{2j\Delta C_0\omega_{0}} & 0 \\ \\
    \dfrac{m}{2j\Delta C_0\omega_{-1}} & \gamma_0(\omega_{0}) + \dfrac{1}{j\Delta C_0\omega_{0}} + & \dfrac{m}{2j\Delta C_0\omega_{1}} \\ \\
    0 & \dfrac{m}{2j\Delta C_0\omega_{0}} & \gamma_0(\omega_{1}) + \dfrac{1}{j\Delta C_0\omega_{1}}
\end{bmatrix}
\end{equation}
\end{widetext}
and
\begin{equation}
\ma{R}=\frac{\eta}{4}\mbox{diag}[k_{-1}S^{(-1)}, k_{0}S^{(0)}, k_{1}S^{(1)}]
\end{equation}
and
$\tilde{\ve{E}}^{i}(\ve{r}_0)=\tilde{\ve{E}}^{i}(\ve{r}_l)$ of Eq.~(\ref{incident field after image}) at $l=0$.

\subsection{The total fields}
The total fields contain different contributions from the impinging and reflected wave by the PEC plate, as well as, of course, by the scattering due to the wires.
Thus,
the total field at the $n$-th harmonic reads
\begin{eqnarray}
&& E_{z,n}=E^i e^{-j(k_x^ix - k_y^iy)} - E^i e^{-j(k_x^ix + k_y^iy)} + \nonumber \\
&& -\frac{\eta k_n}{4} {\cal A}_n \sum_{l=-\infty}^{\infty}\left[ H_{0}^{(2)}\left(k_n\sqrt{(x-la)^2+(y-h)^2}\right) e^{-j\psi_n l}\right. \nonumber \\
&& - \left.H_{0}^{(2)}\left(k_n\sqrt{(x-la)^2+(y+h)^2}\right) e^{-j\psi_n l}\right]
\end{eqnarray}
After applying Poisson summation and separating between the specular reflection term and the higher diffraction harmonics, the total field above $y=h$ are given by
\begin{equation}\label{Total field}
E_{z,n}=E^i e^{-j(k_x^ix - k_y^iy)}\delta_n +
      \sum_{m=-\infty}^{\infty}E^r_{mn}{e^{-j(k_{x,mn} x+\beta_{mn}y)} }
\end{equation}
with
\begin{equation}\label{Emn}
E^r_{mn}=- E^i\delta_n-\frac{j\eta k_n \sin(\beta_{mn} h)}{\beta_{mn}a} {\cal A}_n
\end{equation}
where $\delta_n$ is the delta of Kronecker, $\delta_n=1$  for $n=0$ and $\delta_n=0$ otherwise,  ${\cal A}_n$ is found as a solution to Eq.~(\ref{Eq A}),
and $k_{x,mn}$ and $\beta_{mn}$ are  the $x$ and $y$ components of the propagation wavevector. The former is given by
\begin{equation}
k_{x,mn}=\frac{2\pi m}{a} + \frac{\psi_n}{a}
\end{equation}
whereas the latter by Eq.~(\ref{betam}). Clearly, $k_{x,mn}$ and $\beta_{mn}$ satisfy the free-space dispersion relation $k_{x,mn}^2 + \beta_{mn}^2=k_n^2$.

\subsection{Space-time diffraction orders}
From Eq.~(\ref{Total field}) together with Eq.~(\ref{betam}) it is clear that in the space-time modulated metagrating structure, diffraction lobes are a consequence of the geometrical periodicity $a$ as well as of the spatial and temporal modulation periodicity.
Specifically, propagating diffraction order of spatial order $m$ and temporal order $n$ will exists if
\begin{equation}
\left|\frac{2\pi}{a}\left[m+\frac{n}{3}\right] + k_x^i \right|<\frac{\omega+n\omega_n}{c}
\end{equation}
which boils down to either
\begin{eqnarray}\label{ST Diff Ord}
0\le m+\frac{n}{3} + \tilde{a}\sin\theta^i \le \tilde{a} + n\delta\tilde{a} \nonumber\\
\mbox{or} \nonumber\\
-\tilde{a}-n\delta\tilde{a}\le m+\frac{n}{3}+\tilde{a}\sin\theta^i\le0
\end{eqnarray}
where $\tilde{a}=a/\lambda$ and $\delta=\omega_m/\omega$.
Propagating reflected wave harmonics propagate at angle
%
%
%
\begin{equation}
\theta_{mn}=\sin^{-1}(k_{x,mn}/k_n)
\end{equation}
measured with respect to the $y$ axis, where for reflected waves $\theta$ is defined as positive/negative for wavenumbers in the first/second $xy$ quadrant. Note that for the incoming, impinging wave, the definition is opposite.

Now it is clear that in order to cancel specular reflection at the incident wave frequency it is required to nullify, or at least minimize, the following term that consists of the specular reflection by the PEC plate and the $n=0$ temporal harmonic of the $m=0$ spatial harmonic. That is,
\begin{equation}\label{Cancel Spec Ref}
-E^i - \frac{j\eta k}{a} \frac{\sin(k_y^i h)}{k_y^i} A_0 = 0
\end{equation}
where $A_0$ is given by solving Eq.~(\ref{Eq A}).

\subsection{Energy balance}
The net power flux carried by the $mn$ propagating reflected harmonics through the $y=h$ plane is given by
\begin{equation}\label{Smn}
S_{mn}=\cos\theta_{mn}\frac{|E^r_{mn}|^2}{\eta}.
\end{equation}
In a lossless, passive, and stationary metagratings system it is required that the net outgoing flux will be equal to the net incoming flux by the impinging wave.
In our system, however, due to the parameteric process of the temporal modulation, some additional energy may be pumped into or extracted from the wave system \cite{Hayrapetyanab2013, Hayrapetyan2016, Hayrapetyan2017, Shlivinski2018, Hadad2019}.
Thus, in our system the energy balance reads
\begin{equation}\label{Eneg Bal}
\Delta S_{mod}+ S_{inc}-\sum_{(mn)\in prop} S_{mn}  = 0
\end{equation}
Nonetheless, since the modulation frequency is small compared to the signal frequency, this exceed power is expected to be relatively small, thus $\Delta S_{mod}\ll S_{inc}$.

\section{Nonreciprocal Anomalous Reflection with Near Complete Frequency Conversion}
In this section we use the theory developed above in order to demonstrate one possible functionality of the time-modulated metagratings surface. We consider a space-time modulated metagratings structure with the following parameters: reference frequency $f_r$, reference wavelength $\lambda_r=c/f_r$, inter-wire distance $a=0.6\lambda_r$, wires distance from the PEC plate $h$, wire loading capacitance $C_0=1$pF, loading periodicity $\Delta=0.1\lambda_r$, and wire radius $r_0=0.5$mm.
We introduce space-time modulation of the wires in the lattice such that $\omega_m=0.1\omega_r$ ($\omega_r=2\pi f_r$) with modulation index $m=0.1$, and super-cell size $N=3$.
A stationary metagratings structure, namely, in the absence of the space-time modulation, has been shown to be able to completely nullify the specular reflection \cite{Chalabi2017, Radi2017, Epstein2017, Rabinovich2018, Popov2018, Popov2019, Popov2019b, Epstein2018, Diaz-Rubio2017}. Instead,   the energy is directed to one of the grating lobes of the structure. This functionality has been recently  termed - anomalous reflection. Nonetheless, up to now this effect has been demonstrated only in reciprocal structures, with the exception in \cite{Taravati2019} which explored a metagrating dielectric slab with a continuous space-time modulation of the refractive index. Due to reciprocity, if an impinging wave at $\theta^i$ is fully reflected to $\theta^r$, then a wave arriving at $-\theta^r$ will be reflected with the same efficiency to $-\theta^i$, where $\theta^i$ and $\theta^r$ are measured with respect to the $y$ axis, \emph{and} considered positive when corresponding to waves with positive $k_x$ component of the wave number.
Here, we show that upon space-time modulation this effect can be turned to be strongly non-reciprocal, and moreover, we demonstrate that even with a small  modulation index it is possible to achieve a near  complete  frequency conversion by the temporal modulation.

\begin{figure}[htbp]
  \centering
  \includegraphics[width=80mm]{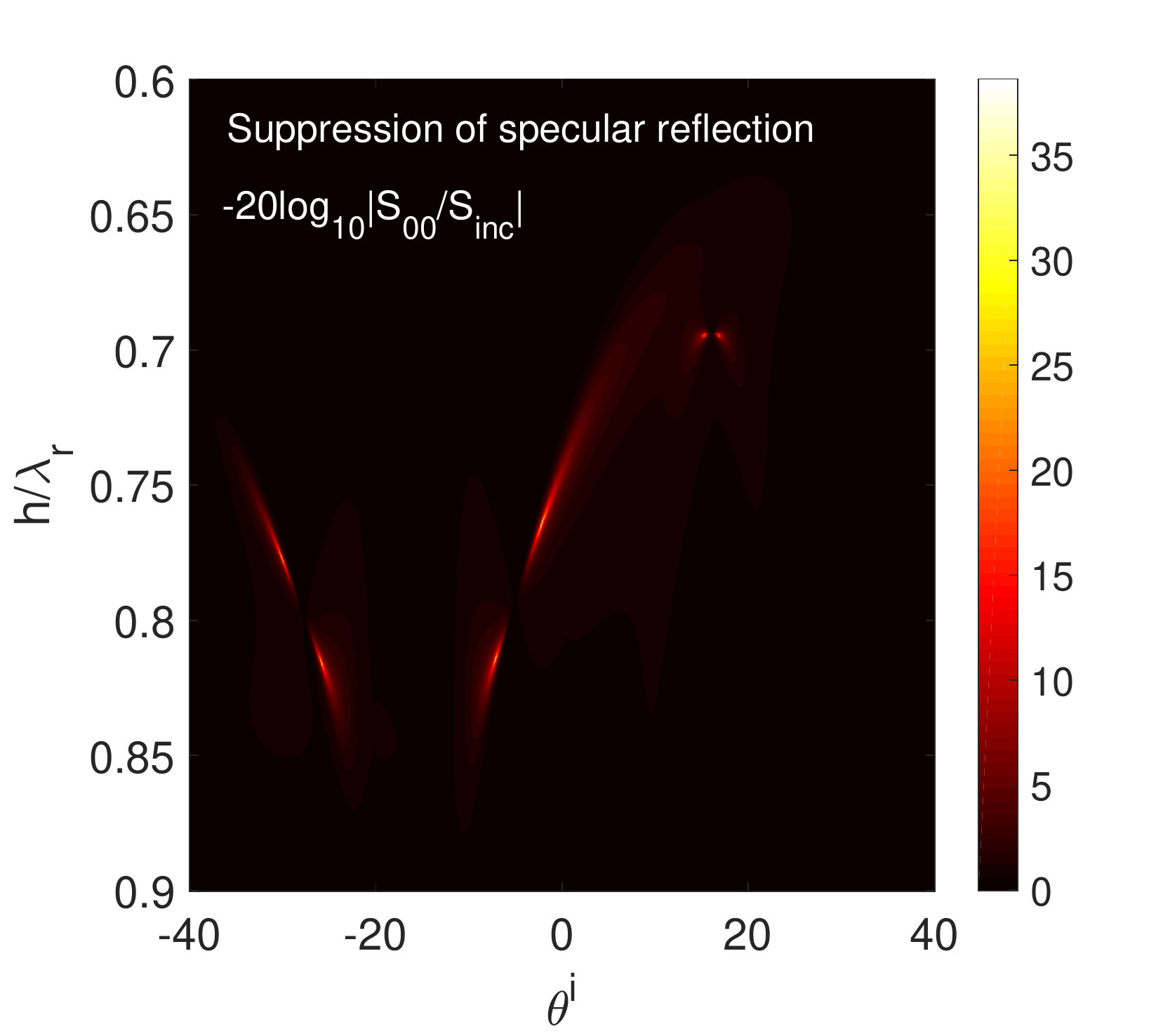}
  \caption{Suppression of the specular reflection [in dB] as function of the incident wave angle $\theta^i$ and the spacing $h$ between the wire-lattice and the PEC plate. This map is calculated for the parameters that are listed in the text, specifically, $a/\lambda=0.6$. In the map, the \emph{dark areas correspond to full specular reflection}, whereas significant suppression of specular reflection is characterized by light colors. In these regions where specular reflection is prohibited, significant frequency conversion takes place into another space-time diffraction order as shown the Fig.~\ref{fig3} below.}
  \label{fig2}
\end{figure}

With the parameters listed above, in Fig.~\ref{fig2} we show the suppression of the specular reflection as a function of the incidence angle $\theta^i$ and the distance between the lattice and the PEC plate $h$.  
The figure shows the specular reflection intensity, color-coded in logarithmic scale.  For large range of parameters nearly full reflection takes place. However, at certain regions in the reflection intensity map inter-harmonic resonances take place leading to full suppression on the specular reflection.  From the asymmetry in the figure, it is clear that this effect is non-reciprocal, leading to a very different power transmission for impinging waves that are coming from complementary direction, $\theta^i$ or $-\theta^i$.
Let us consider a specific example, with $h=0.7775\lambda_r$ and incidence angle $\theta^i=-30^\circ$.
With these parameters, using Eq.~(\ref{ST Diff Ord}) it is clear that besides the specular reflection (fundamental harmonic $(m,n)=(0,0)$), also the higher order $(m,n)=(1,-1)$ and $(m,n)=(0,1)$ harmonics are propagating. This means, that the incident wave power that is not specularly reflected has to efficiently couple to one or two of these additional propagating diffraction orders.
This is demonstrated in Fig.~\ref{fig3} below. In the figure, the power flux density along the $y$ direction, $S_{mn}$, is shown for the propagating harmonics. The specular reflection wave due to an incident wave at $\theta^i=-30^\circ$ is shown in blue. At the desired reference frequency, specular reflection is completely suppressed. The incident wave energy experiences practically complete conversion to the $(m,n)=(1,-1)$ harmonic that propagates at frequency $\omega_{-1}=\omega_r-\omega_m$ which is reflected towards $\theta_{(1,-1)}\approx 42^\circ$. As opposed to that, the $(m,n)=(0,1)$ harmonic which is also propagating is very weakly excited. On the contrary, in the absence of inter-harmonic resonances as shown in Fig.~\ref{fig2}, the wave that impinges at the complementary direction, i.e., at $\theta^i=30^\circ$ experiences nearly complete specular reflection, and therefore, practically, no additional propagating harmonics will be excited.

\begin{figure}
  \includegraphics[width=80mm]{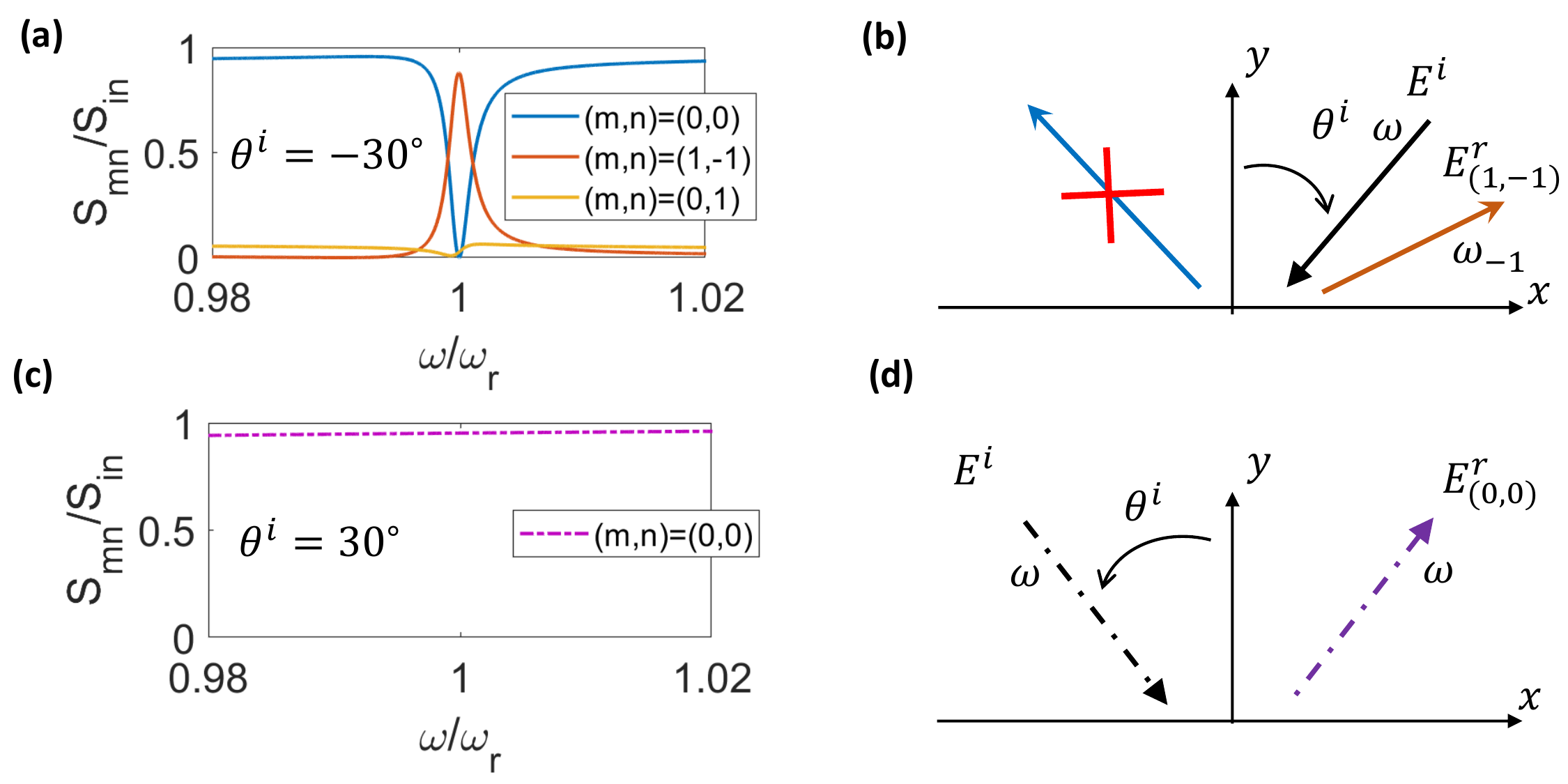}
  \caption{Non-reciprocal scattering and high efficient frequency conversion. (a) With $\theta^i=-30^\circ$. In this case there is no specular reflection at $\omega=\omega_r$ as shown by the continuous blue line. The dominant reflected wave corresponds to the $(m,n)=(1,-1)$ space time harmonics which propagates  at angle $\theta^r_{1,-1}\approx 42^\circ$ and at frequency $\omega_{-1}=\omega-\omega_m$. Due to the parametric modulation some scattering to additional harmonics may take place but with very low efficiencies. For example,  an additional propagating harmonics $(m,n)=(0,1)$ shown in yellow is practically not excited. An illustration of the overall response in this case is shown in (b). As opposed to that, with $\theta^i=30^\circ$ practically regular specular reflection takes place as shown in (c), with $\theta^r=30^\circ$. In this case the metagrating surface practically behaves as a simple reflector, as illustrated in (d). Yet, additional scattering harmonics reduce the overall efficiency by some minor extent. }
  \label{fig3}
\end{figure}
%
%

\section{Some Practical Considerations and Conclusions}
A few words on practical realization of the proposed design. In the example considered above only $N=3$ modulation regions are required per unit cell. The distance between the wires is large on the wavelength thus enabling a convenient wiring system. Furthermore, we note that all the capacitors that are distributed on each wire can be modulated simultaneously by modulation voltage that is applied at the two ends of each wire.   Moreover, an additional simplification of the practical implementation may be achieved if each family of wires that are subject to the same modulation (namely, all the wires of the same color in Fig.~\ref{fig1}) will be placed at a different height with respect to the ground plane. In that case, the modulation scheme may be designed using even a smaller number of modulation drivers. Note though that the analysis carried out in this paper should be slightly adjusted for this particular case.

To conclude, in this paper we have developed a theoretical model based on the discrete-dipole-approximation and polarizability theory for the scattering by a two-dimensional lattice of space-time modulated resonant capacitively loaded wires. Using this simple methodology we demonstrate the design of a significantly non-reciprocal anomalous reflection using metagrating structure with \emph{electrically large} unit cells that contain a single wire each. Thus, significantly simplifying the practical requirements for the modulation system and hence opening a realistic possibility for actual fabrication of such devices.
The non-reciprocal scattering process described here is also associated with efficient frequency conversion of the anomalous refracted beam. While the model explored in this paper involves capacitively loaded wires, the same approach may be also augmented to the optical frequency regime using fast modulation techniques \cite{Phare2015, Liu2011}. And with possible implications in radio-frequency devices as well as for more efficient photovoltaic processes \cite{Green2012, Zhu2014}.
Moreover, we note that the model developed in this paper may be readily augmented to solve the excitation response due to a localized source by taking an approach akin to \cite{Hadad2011,Hadad2013_2}.

\begin{acknowledgments}
Y.~Hadad would like to acknowledge support by the Alon Fellowship granted by the Israeli
Council of Higher Education (2018-2021).
\end{acknowledgments}


\begin{thebibliography}{1}
\bibitem{Comm}
In the pollowing we apply the proposed approach on infinite wires excited by electromagnetic wave polarized along the wire axis. While the theory remains essentially the same, in this context a better term would be susceptability rather than polarizability.
\end{thebibliography}
\end{document}